\documentclass[twocolumn,aps,prb,showpacs,floatfix,superscriptaddress]{revtex4}
\usepackage[latin9]{inputenc}
\usepackage{bm}
\usepackage{amsmath}
\usepackage{amssymb}
\usepackage{graphicx}

\makeatletter
\@ifundefined{textcolor}{}
{%
 \definecolor{BLACK}{gray}{0}
 \definecolor{WHITE}{gray}{1}
 \definecolor{RED}{rgb}{1,0,0}
 \definecolor{GREEN}{rgb}{0,1,0}
 \definecolor{BLUE}{rgb}{0,0,1}
 \definecolor{CYAN}{cmyk}{1,0,0,0}
 \definecolor{MAGENTA}{cmyk}{0,1,0,0}
 \definecolor{YELLOW}{cmyk}{0,0,1,0}
}

\usepackage{bm}

\newcommand{\Ca}{CaFe$_2$As$_2$}
\newcommand{\Ba}{BaFe$_2$As$_2$}

\newcommand{\mat}[1]{\hat{\bm{\mathrm{#1}}}}

\makeatother

\begin{document}

\title{In-plane uniaxial stress effects on the structural and electronic
properties of {\Ba} and {\Ca}}

\author{Milan Tomi{\'{c}}}

\affiliation{Institut f\"ur Theoretische Physik, Goethe-Universit\"at Frankfurt, Max-von-Laue-Stra{\ss}e
1, 60438 Frankfurt am Main, Germany}

\author{Harald O. Jeschke}

\affiliation{Institut f\"ur Theoretische Physik, Goethe-Universit\"at Frankfurt, Max-von-Laue-Stra{\ss}e
1, 60438 Frankfurt am Main, Germany}

\author{Rafael M. Fernandes}

\affiliation{School of Physics and Astronomy, University of Minnesota, Minneapolis,
MN 55455, USA}

\author{Roser Valent{\'\i}}

\affiliation{Institut f\"ur Theoretische Physik, Goethe-Universit\"at Frankfurt, Max-von-Laue-Stra{\ss}e
1, 60438 Frankfurt am Main, Germany}

\date{\today}
\begin{abstract}
Starting from the orthorhombic magnetically ordered phase, we investigate
the effects of uniaxial tensile and compressive stresses along \textbf{a},
\textbf{b}, and the diagonal \textbf{a}+\textbf{b} directions in {\Ba}
and {\Ca} in the framework of \textit{ab initio} density functional
theory (DFT) and a phenomonological Ginzburg-Landau model. While --contrary
to the application of hydrostatic or $c$-axis uniaxial pressure--
both systems remain in the orthorhombic phase with a pressure-dependent
nonzero magnetic moment, we observe a sign-changing jump in the orthorhombicity
at a critical uniaxial pressure, accompanied by a reversal of the
orbital occupancy and a switch between the ferromagnetic and antiferromagnetic
directions. Our Ginzburg-Landau analysis reveals that this behavior
is a direct consequence of the competition between the intrinsic magneto-elastic
coupling of the system and the applied compressive stress, which helps
the system to overcome the energy barrier between the two possible
magneto-elastic ground states. Our results shed light on the mechanisms
involved in the detwinning process of an orthorhombic iron-pnictide
crystal and on the changes in the magnetic properties of a
system under uniaxial stress. % by 40K/GPa both in {\Ba} and {\Ca}.

\end{abstract}

\pacs{74.70.Xa,61.50.Ks,71.15.Mb,64.70.K-}

\maketitle
%62.50.-p 	High-pressure effects in solids and liquids

%64.70.K- 	Solid-solid transitions

%68.35.Rh 	Phase transitions and critical phenomena

%71.15.Mb 	Density functional theory, local density approximation, gradient and other corrections

%71.15.Pd 	Molecular dynamics calculations (Car-Parrinello) and other numerical simulations

%74.70.Xa 	Pnictides and chalcogenides

\section{Introduction}

Since the discovery of high-$T_{{\rm c}}$ superconductivity in Fe-based
materials in 2008,~\cite{Kamihara2008} an enormous amount of effort
has been invested to understand the microscopic behavior of these
systems. Iron pnictides and chalcogenides become superconductors either
by hole- or electron-doping the systems, by application of external
pressure or by a combination of both. In particular, uniaxial pressure
is currently being intensively discussed as a possible route towards
modifying the structural, magnetic and even superconducting properties
of these systems. A regular sample below its magnetic and structural
transition temperatures displays an equal number of opposite twin
orthorhombic domains, effectively canceling out its anisotropic properties.
To circumvent this issue and obtain a single orthorhombic domain sample,
uniaxial tensile stress has been widely employed to detwin iron pnictides
like {\Ba} and {\Ca} \cite{Chu2010,Tanatar2010,Kuo2011,Dhital2012,Liang2011,Blomberg2012,Fisher12}
and unveil its anisotropic properties -- which have been argued to
originate from electronic nematic degrees of freedom.~\cite{Fang08,Xu08,FernandesPRL10}
Theoretically, although it is clear that in the tetragonal phase the
applied uniaxial pressure acts as a conjugate field to the orthorhombic
order parameter, condensing a single domain,~\cite{Fernandes11}
the nature of the detwinning process deep inside the orthorhombic
phase remains an open question, since different mechanisms might be
at play -- such as twin boundary motion or reversal of the order parameter
inside the domains.~\cite{Blomberg11,Fisher12}

Besides promoting detwinning, uniaxial strain has also been shown
to affect the thermodynamic properties of the iron pnictides. Recent
neutron scattering experiments on {\Ba} under compressive stress
along the in-plane \textbf{b} direction reported a progressive shift
to higher temperatures of the magnetic transition~\cite{Dhital2012}
- a behavior also seen in $\mathrm{BaFe_{2}\left(As_{1-x}P_{x}\right)_{2}}$
by thermodynamic measurements.~\cite{Kuo12} -- and an apparent
reduction of the magnetic moment~\cite{Dhital2012}. Moreover, Blomberg
\textit{et al.} observed a significant uniaxial structural distortion
in {\Ba} under tensile stress, suggesting an enhanced response
to external strain.~\cite{Blomberg2012} More recently, it was found
that epitaxially strained thin films of FeSe on a SrTiO$_{3}$ substrate
show an increase in critical superconducting temperatures up to 65
K, the highest reported $T_{{\rm c}}$ so far.~\cite{Wang2012} Clearly,
crystal lattice strain plays a key role for the magnetic, structural
and superconducting properties in Fe-based superconductors and a better
understanding of the microscopic origin of such behavior is desirable.

In this work we combine density functional theory (DFT) calculations
and Ginzburg-Landau phenomenology to analyze the effects of uniaxial
compressive stress as well as uniaxial tensile stress on the magnetic,
electronic and structural properties of {\Ba} and {\Ca} at low
temperatures, deep inside the ordered phase. Stress is measured in
terms of equivalent hydrostatic pressure, $P=\mathrm{Tr}(\mat{\sigma})/3$,
where $\mat{\sigma}$ is the stress tensor matrix and positive and
negative pressures correspond to applying compressive and tensile
stresses respectively. Our \textit{ab initio}-derived estimates for
the elastic constants in the orthorhombic phase agree well with experimental
values. While no sign of a true structural or magnetic phase transition
is observed in the range of pressures between $-2$~GPa and $2$~GPa,
at a critical pressure we observe a reversal of the magnetization,
\textit{i.e.} exchange of ferromagnetic (FM) and antiferromagnetic
(AFM) directions, simultaneous to a discontinuous change in the orthorhombic
order parameter $a-b$, which also changes sign. This behavior has
important consequences on the orbital $d_{xz}$ and $d_{yz}$ occupancies
and is also related to the shift of the magnetic ordering temperature,
as we argue below.

Furthermore, by employing a phenomenological Ginzburg-Landau model,
we show that this behavior is intimately connected to the magneto-elastic
coupling of the system, which by itself acts as an intrinsic conjugate
field to the orthorhombic order parameter. As the applied compressive
stress is enhanced towards a critical value, it eventually overcomes
the effects of the magneto-elastic coupling, rendering the zero-pressure
state energetically unstable and resulting in a simultaneous reversal
of the magnetization and the orthorhombic order parameter. Comparison
of the DFT-derived critical uniaxial pressures for {\Ca} and {\Ba},
combined with the Ginzburg-Landau result that the critical pressure
is proportional to the magneto-elastic coupling, suggests that the
latter is larger in {\Ca} than in {\Ba}. We also propose low-temperature
detwinning measurements to compare the experimental critical pressure
with our \textit{ab initio} estimates in order to clarify the dominant
mechanism behind the detwinning process of orthorhombic iron pnictide
crystals.

The paper is organized as follows: in Section II we summarize our
computational methods, and in Section III we present our DFT results.
Section IV is devoted to the Ginzburg-Landau analysis and comparison
with DFT calculations. Section V contains the discussions and concluding
remarks. Details about the \textit{ab initio} method are presented
in the Appendix.

\section{Computational Methods}

Electronic structure calculations were performed within DFT with the
Vienna ab initio simulations package (VASP)~\cite{Kresse1993} with
the projector augmented wave (PAW) basis~\cite{Bloechl1994} in the
generalized gradient approximation (GGA). All our structural relaxations
were performed under constant stress using the Fast Inertial Relaxation
Engine (FIRE).~\cite{Bitzek2006,Tomic2012} For this purpose, we
had to modify the algorithm accordingly (see Appendix A). Every 10
steps, we cycled through non-magnetic, ferromagnetic, antiferromagnetic-checkerboard,
stripe-type antiferromagnetic (along unit cell axis \textbf{a}) and
stripe-type antiferromagnetic (along unit cell axis \textbf{b}) spin
configurations, and then we continued the relaxation with the lowest
energy spin configuration. As a final converged magnetic configuration
in the orthorhombic phase we always found stripe-type antiferromagnetic
order either along \textbf{a} or along \textbf{b}, as discussed below.

The energy cutoff in the calculations was set to 300~eV and a Monkhorst-Pack
uniform grid of $(6\times6\times6)$ points was used for the integration
of the irreducible Brillouin zone (BZ).

\section{Results and Discussion}

\subsection{BaFe$_{2}$As$_{2}$}

\begin{figure*}[bth]
  \includegraphics[width=1\textwidth]{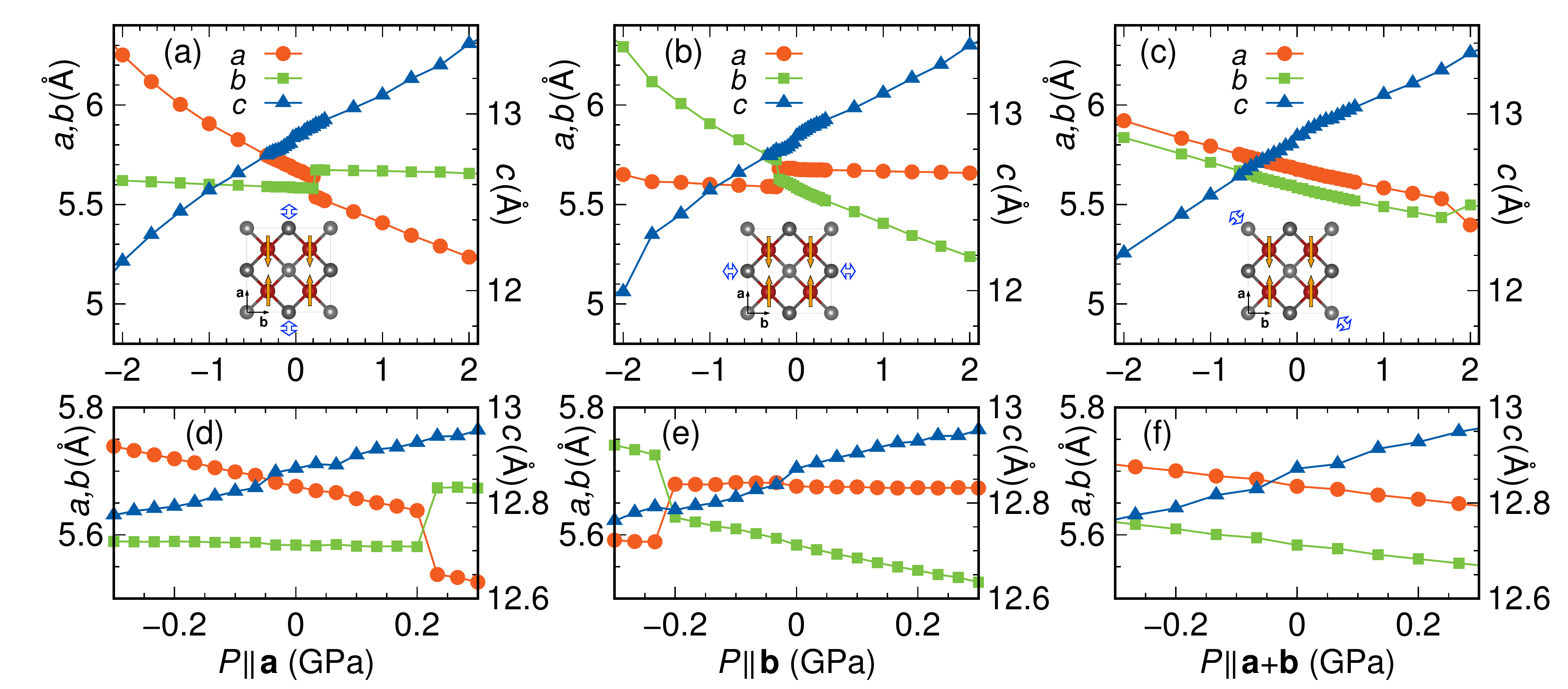} \caption{(Color
    online) Evolution of the unit cell parameters in {\Ba} under
    application of uniaxial stress in the equivalent hydrostatic
    pressure range $[-2\,{\rm GPa},2\,{\rm GPa}]$ (a) along
    \textbf{a}, (b) along \textbf{b} and (c) along
    \textbf{a}+\textbf{b}. Panels (d)-(f) show the corresponding zoom
    of the pressure dependence of the lattice parameters in the range
    $[-0.3\,{\rm GPa},0.3\,{\rm GPa}]$. Negative pressures correspond
    to tensile stress while positive pressures correspond to
    compressive stress. Note, that the relationship between axes and
    iron moments shown in the inset in (a) is valid for
    $P\parallel{\bf a}<0.22$~GPa, in (b) for $P\parallel{\bf
      b}>-0.22$~GPa. For a discussion of the reversal of AFM order,
    see the text. \label{Ba_latticeparameters} }
\end{figure*}

Starting from the low-temperature orthorhombic structure with stripe
magnetic order, we performed structure relaxations under applied uniaxial
tensile and compressive stresses along \textbf{a} (AFM direction),
\textbf{b} (FM direction) and the plane-diagonal \textbf{a}+\textbf{b}
direction for both {\Ba} and {\Ca} (see inset of Figure~\ref{Ba_latticeparameters}~(a)).
We measure stress in units of the equivalent hydrostatic pressure,
$P=\mathrm{Tr}(\mat{\sigma})/3$, with $\mat{\sigma}$ denoting the
stress tensor matrix. We simulated pressures in the range between
$-3$~GPa and $3$~GPa. In the tensile stress range, below $-2.7$~GPa
we observe in both systems that, for stress along \textbf{a}, a sudden
expansion in $a$ and contraction in $b$ and $c$ axes occurs. A
similar situation arises when pulling apart along \textbf{b}. This
feature signals the extreme case of absence of bonding within the
material, and for this reason this pressure range will be excluded
from further discussion.

Figure~\ref{Ba_latticeparameters} shows the evolution of lattice
parameters for {\Ba} as a function of uniaxial stress along \textbf{a},
\textbf{b} and \textbf{a}+\textbf{b}. We consider both compressive
stress (positive pressure) and tensile stress (negative pressure).
At $P=0$~GPa, we have $a$ (AFM direction) $>$ $b$ (FM direction).
{\Ba} remains in the orthorhombic phase with nonzero increasing
magnetic moment for large tensile stress (negative pressure). Pulling
apart (\textit{i.e.} $P<0$) along the (longer) AFM direction \textbf{a}
(Figure~\ref{Ba_latticeparameters}~(a)) the system expands along
\textbf{a}, strongly compresses along \textbf{c} and shows almost
no changes along \textbf{b}; similarly, pulling apart along the (shorter)
FM direction \textbf{b} (Figure~\ref{Ba_latticeparameters}~(b))
$b$ expands, $c$ compresses and $a$ shows almost no changes except
at the pressure $P=-0.22$~GPa (Figure~\ref{Ba_latticeparameters}~(e)).
At this point, {\Ba} shows a sudden jump in the orthorhombicity
where \textbf{a} becomes the shorter axis and \textbf{b} becomes the
longer axis. This interchange happens with a rotation of the magnetic
order by 90 degrees, \textit{i.e.} the FM direction becomes parallel
to the \textbf{a} axis while the AFM direction becomes parallel to
the \textbf{b} axis. We will discuss this feature further below. Note
that tensile stress along \textbf{a}+\textbf{b} acts similarly on
both \textbf{a} and \textbf{b} directions, which expand, while the
\textbf{c} direction strongly compresses (Figures~\ref{Ba_latticeparameters}~(c)
and (f)).

Under application of compressive stress (positive pressure), we observe
in all three cases a strong expansion along \textbf{c} and a compression
along the direction of applied stress\textbf{ }(\textbf{a}, \textbf{b}
or \textbf{a}+\textbf{b}). For the cases where pressure is applied
along $\mathbf{a}$ or $\mathbf{b}$, we observe almost no changes
or a slight expansion along \textbf{b} and \textbf{a}, respectively.
%RMF removing this part A slight expansion along a has also been measured upon application of uniaxial stress along b by Dhital et al..  remains in all cases orthorhombic with nonzero decreasing magnetic moment.
 Since $a>b$ at zero stress, we observe the inversion of axes followed
by a jump in orthorhombicity and a 90 degree rotation of the magnetization
when stress is applied along \textbf{a} at $P=0.22$~GPa (Figures~\ref{Ba_latticeparameters}~(a)
and (d)). This inversion of axes, with $b>a$ for all higher compressive
stress values means that the spin configuration shown in the inset
of Figure~\ref{Ba_latticeparameters}~(a) should now be turned by
90 degrees, with \textbf{b} pointing along the AFM direction. Such
an inversion is also observed for compressive stress along \textbf{a}+\textbf{b}
at much larger pressures of $P=2$~GPa.

\begin{figure}
  \includegraphics[width=0.5\textwidth]{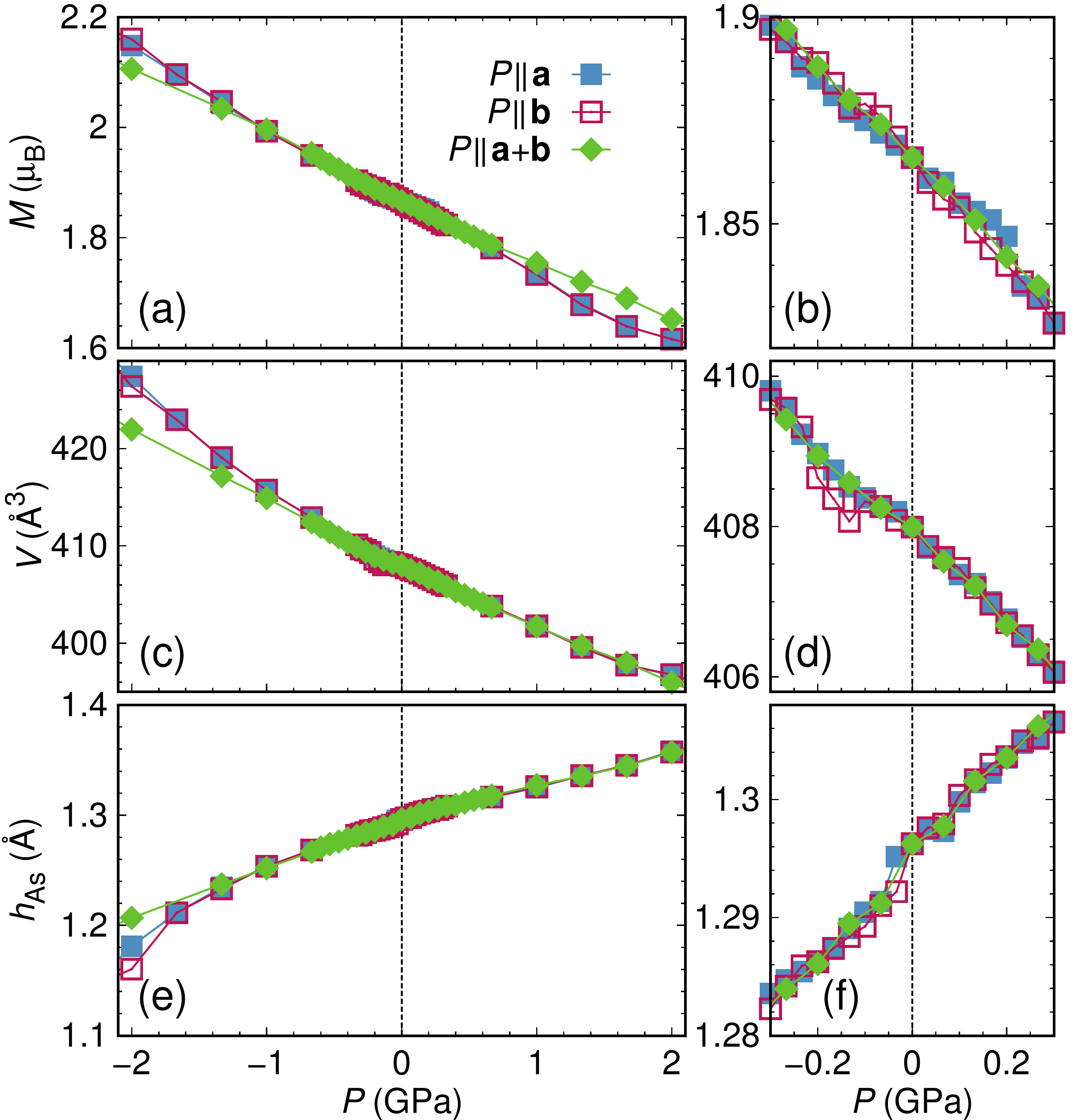} \caption{(Color
    online) (a) Evolution of the magnetic moments of iron, (c) of the
    unit cell volume and (e) of the pnictogen height under uniaxial
    pressure in the range $[-2\,{\rm GPa},2\,{\rm GPa}]$. Panels (b),
    (d) and (e) show the corresponding zoom of the pressure dependence
    of these quantities in the range $[-0.3\,{\rm GPa},0.3\,{\rm
      GPa}]$.  Negative pressures correspond to tensile stress while
    positive pressures correspond to compressive
    stress. \label{ba122moments} }
\end{figure}

Figure~\ref{ba122moments} shows the evolution of magnetic moment,
volume and As height in {\Ba} as a function of stress. The three
quantities show a clearly monotonic behavior independent of the
applied stress direction except for small jumps at the pressures
$P=-0.22$~GPa (for stress along \textbf{a}) and $P=0.22$~GPa (for
stress along \textbf{b}) where the tetragonal condition is almost
fulfilled ($a$ $\approx$ $b$) (Figures~\ref{ba122moments}~(b), (d),
(f)).  We also note here how magnetic moments in {\Ba} respond to different direction
of pressure application. The highest rate of suppression, of roughly 
0.1$\mu_B$/GPa is achieved when pressure is applied within the $ab$-plane,
while application of pressure along the \textbf{c}-axis actually results
in magnetic moment increase by 0.03$\mu_B$/GPa~\cite{Tomic2012}.  Even though DFT calculations
overestimate the value of the ordered Fe magnetic moment at $P=0$~GPa,
it is to be expected that the relative changes in magnetic moment
should provide a reliable description of the situation of {\Ba} under
pressure as shown in previous
studies.~\cite{Zhang2009,Colonna2011a,Colonna2011b,Tomic2012}

Except for the pressures $P=-0.22$~GPa (for stress along \textbf{a})
and $P=0.22$~GPa (for stress along \textbf{b}), stress always enforces
a certain degree of orthorhombicity and the system remains magnetically
ordered with a decreasing ordered moment as a function of compressive
stress (Figure~\ref{ba122moments}~(a)). Moreover, since the $c$
axis continually expands from negative to positive pressures, $h_{As}$
increases accordingly as a function of stress (Figure~\ref{ba122moments}~(e)).
These features have a direct consequence on the electronic properties
of the system.

\begin{figure*}[bth]
  \includegraphics[width=1\textwidth]{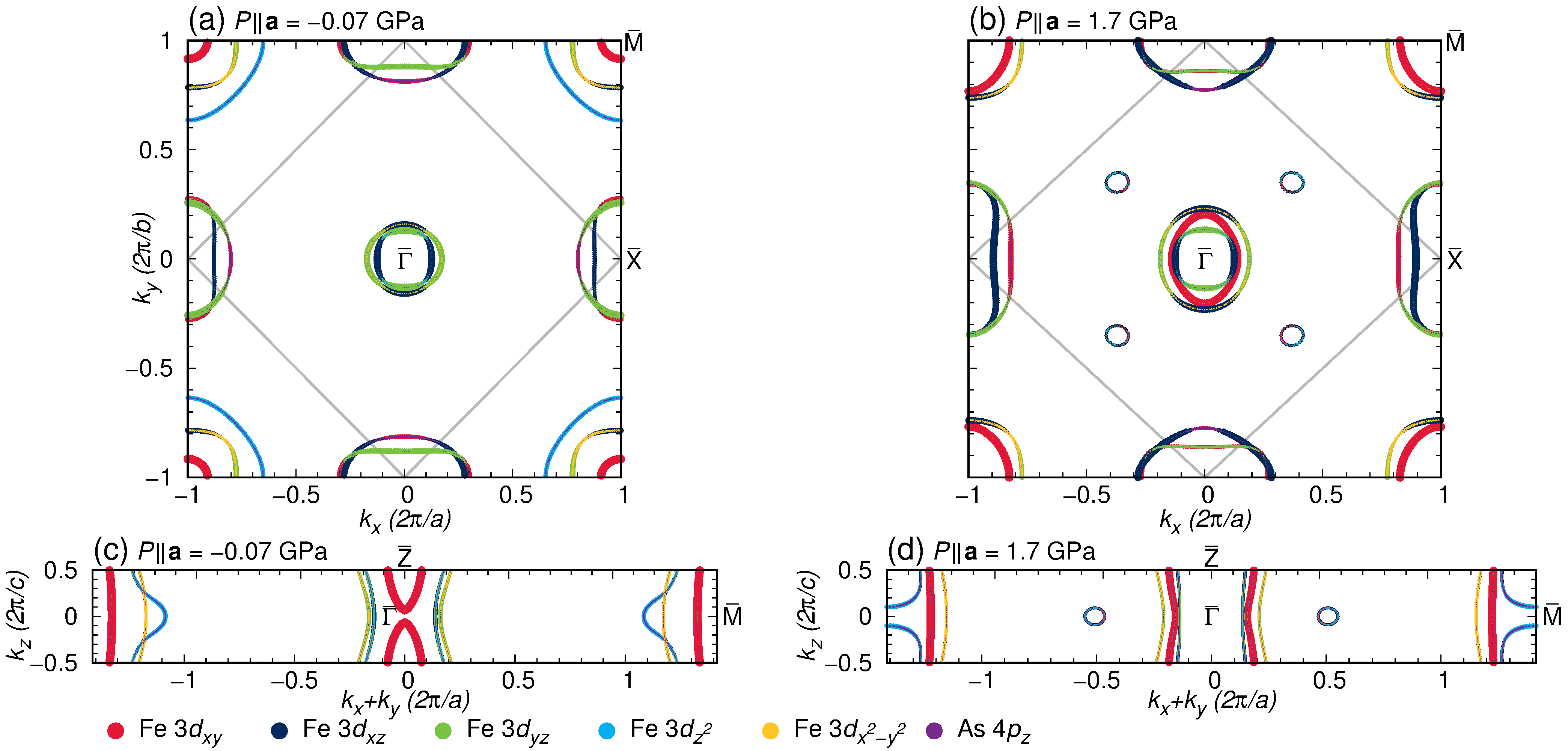} \caption{(Color
    online)  Fermi surface of {\Ba} for two pressure values of uniaxial
    stress applied along the \textbf{a} axis shown in the 1Fe/unit
    cell equivalent BZ (see Ref.~\protect\onlinecite{Boeri} for the BZ
    path definition). Panels (a) and (b) show $k_{z}=0$ cuts of the
    Fermi surface at pressures of -0.07~GPa and 1.7~GPa respectively,
    while panels (c) and (d) show vertical cuts along the diagonal of
    the BZ for pressures of -0.07~GPa and 1.7~GPa. Grey lines on
    panels (a) and (b) denote boundaries of the 2 Fe/unit cell
    BZ. \label{ba_fs} }
\end{figure*}

As an illustration, we show in Figure~\ref{ba_fs} 
the (non-spin polarized) Fermi surface
of {\Ba} under application of uniaxial stress $P=-0.07$~GPa and
$P=1.7$~GPa applied along \textbf{a}  in
the 1Fe/unit cell equivalent Brillouin zone. We would like to note
that correlation effects beyond DFT as implemented in DFT+DMFT (dynamical
mean field theory), which are known to give a good agreement between
the calculated Fermi surfaces and angle-resolved photoemission measurements
in the Fe pnictides,~\cite{Aichhorn2009,Yin2011,Aichhorn2011,Ferber2012a,Ferber2012b,Werner2012}
have not been included here. %\textbf{\emph{(this is not reconstructed by AFM order? We should explain
%this point)}}. 
%At ambient pressure conditions (Figure~\ref{ba_fs}
%(b)) we observe three hole pockets centered around $\bar{\Gamma}$
%and $\bar{M}$ and two electron pockets centered around $\bar{X}$
%\textbf{\emph{(this panel of the figure is missing)}}.
Modest tensile stress of 0.07~GPa leads to the disappearance of the
$3d_{xy}$ hole pocket around $\bar{\Gamma}$ in the $k_{z}=0$ plane
(see Figure~5~(a) in Ref.~\onlinecite{Tomic2012}). On the other
hand, when compressive stress is applied, the hole pockets around
$\bar{\Gamma}$ significantly change in size, and additionally at
a pressure of 1.7~GPa, small electron pockets, of majority $3d_{xy}$
and $3d_{z^{2}}$ character, appear along the $\bar{\Gamma}-\bar{M}$
directions of the BZ (Figures~\ref{ba_fs}~(c) and (e)). The increase
of the $3d_{xy}$ hole pocket size with increasing uniaxial stress
can be explained by the reduction of Fe-Fe distances along the \textbf{a}
axis, leading to an increased contribution of $3d_{xy}$ - $3d_{xy}$
bonding. In fact, the effects of tensile and compressive stress on
the electronic structure shown for the example of stress along \textbf{a}
can be seen also in our calculations for both stress along \textbf{b}
and along \textbf{a}+\textbf{b}.

\begin{figure}[bth]
\includegraphics[width=0.45\textwidth]{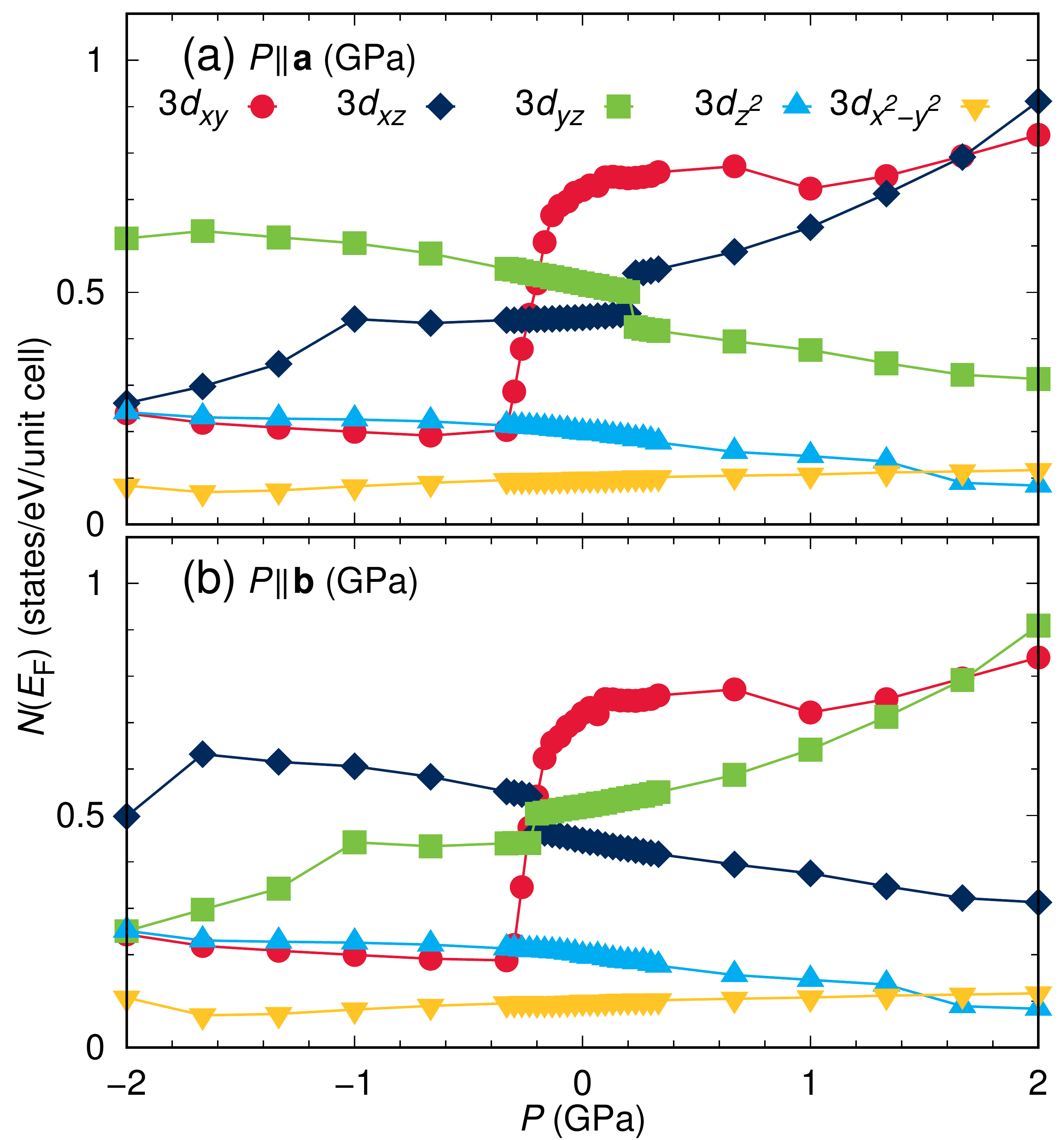} \caption{(Color online) Evolution of the orbital resolved density of states
of {\Ba} at the Fermi level $N(E_{{\rm F}})$ with stress (a) applied
along \textbf{a}, and (b) along \textbf{b}. Lines joining the calculated
points are a guide for the eye. \label{ba122dos} }
\end{figure}

%A more detailed information of the orbital occupation can be obtained
In Figure~\ref{ba122dos} we analyze the orbitally-resolved density
of states at the Fermi level $N(E_{{\rm F}})$. Applying stress both
along \textbf{a} and \textbf{b} has the same effect on the total density
of states of both {\Ba} and {\Ca}, but there is a selective
orbital order as shown in Figure~\ref{ba122dos}. $N(E_{{\rm F}})$
is predominantly of $3d_{xz}$ character when $a>b$ and of $3d_{yz}$
character when $a<b$. This means that the dominant character switches
from $3d_{yz}$ to $3d_{xz}$ at $\mat{\sigma}\parallel{\bf a}\approx0.22$~GPa,
and from $3d_{xz}$ to $3d_{yz}$ at $\mat{\sigma}\parallel{\bf b}\approx-0.22$~GPa
as expected.

%Our analysis of the data at the
%pressures where the inversion of axes occurs allows us to obtain estimates
%for the range of values some key parameters of phenomenological model 
%given in Ref.~\onlinecite{Cano2012} can take.

%(Maybe something can be said about small electron pockets)

%Even though our simulations don't allow calculations in the
%superconducting phase, we can indirectly speculate that
%superconductivity with $T$ should be reached under strain
%conditions that favor the tetragonal condition.

\subsection{ {\Ca}}

\begin{figure*}[bth]
\includegraphics[width=1\textwidth]{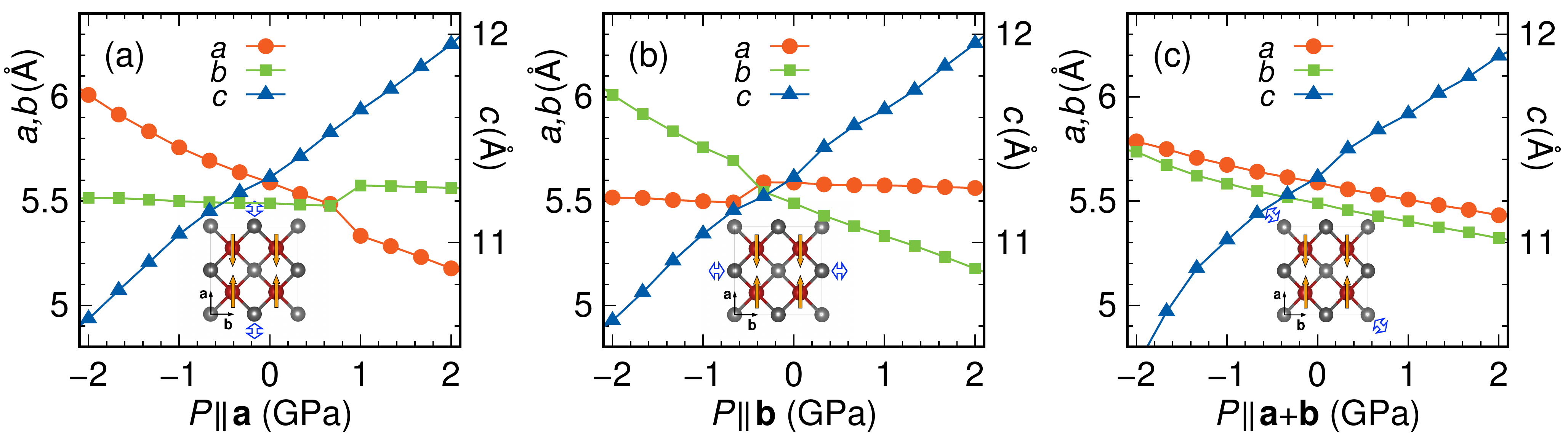} \caption{(Color online) Evolution of the unit cell parameters in {\Ca} under
the application of uniaxial stress in the range $[-2\,{\rm GPa},2\,{\rm GPa}]$
(a) along \textbf{a}, (b) along \textbf{b} and (c) along \textbf{a}+\textbf{b}.
Negative pressures correspond to tensile stress while positive pressures
correspond to compressive stress. Note, that the relationship between
axes and iron moments shown in the inset in (a) is valid for $P\parallel{\bf a}<1$~GPa,
in (b) for $P\parallel{\bf b}>-0.6$~GPa. \label{ca122latticeparameters}}
\end{figure*}

The lattice parameters of {\Ca} under application of compressive
stress along \textbf{a}, \textbf{b} and \textbf{a}+\textbf{b} directions
show a similar overall behavior compared to {\Ba} (see Figure~\ref{ca122latticeparameters})
except for some important shifts of the pressures at which the system
exchanges the FM and AFM directions. When stress is applied along
the \textbf{a} direction, we observe at $P=0.67$~GPa a jump in the
orthorhombic order parameter, with a sign-change, accompanied by a
reversal of the magnetic AFM and FM directions. However, analogously
to the case of {\Ba}, this is not followed by a suppression of
the magnetic moments of iron. In fact, the $c$ axis expands with
applied stress and at $P=0.67$~GPa the $c$ lattice parameter in
{\Ca} is too large for the formation of an interlayer As-As covalent
bond, necessary for a transition to a collapsed tetragonal phase and
suppression of magnetic moments as observed under hydrostatic or $c$-axis
uniaxial pressure.~\cite{Kreyssig2008,Zhang2009,Colonna2011a,Tomic2012}
For tensile stress along the (shorter) \textbf{b} direction, the reversal
of AFM and FM directions happens at $P=-0.33$~GPa followed by a
jump in the orthorhombicity. The magnetic response of {\Ca} is highly
anisotropic as well, but contrary to the case of {\Ba}, the magnetic 
moments in {\Ca} are most effectively suppressed when pressure is
applied along \textbf{c}. We measure  a rate of about 0.1$\mu_B$/GPa~\cite{Tomic2012}. 
 Application of pressure within the $ab$-plane results in
a suppression of the magnetic moment at a rate of 0.02$\mu_B$/GPa.

% In order to understand why {\Ca} shows a jump in orthorhombicity and
% exchanges FM by AFM directions at larger pressures than {\Ba}, we
% estimated the magnetoelastic coupling in both systems within DFT.  In
% {\Ca} we observe that magnetism couples stronger to the lattice
% degrees of freedom than in {\Ba} and therefore larger strains are
% needed in order to have a change in the magnetic ordering.

% When pressure is applied along the diagonal of the $ab$-plane of the
% unit cell, orthorhombicity is preserved in the entire range of
% pressures considered.  $a$ and $b$ undergo a monotonic compression
% with pressure increase, while $c$ expands at the same time. At a
% strain of 2~GPa % is 0.67~GPa meant here?
% {\Ca} undergoes a weak phase transition, bearing traits
% of the orthorhombic to collapsed tetragonal phase transition observed
% under hydrostatic and $c$-uniaxial pressure conditions.

In order to investigate the possibility of a structural and/or magnetic
phase transition at higher pressures, we concentrate now on compressive
stress along the diagonal of the $ab$-plane. We find that orthorhombicity
is preserved up to 7.7~GPa, where a sharp transition to a tetragonal
phase appears. This transition is of first-order type like the orthorhombic
to collapsed tetragonal phase transition under application of hydrostatic
or uniaxial pressure along the \textbf{c} axis~\cite{Tomic2012}
but in this case, changes of magnetic and structural properties take
opposite directions; the \textbf{c} axis undergoes a sudden expansion
of about 9.5\%, and \textbf{a} and \textbf{b} axes contract while
the iron magnetic moments order ferromagneticaly and sharply increase
in value by around 25\%. Interestingly though, contrary to the application
of hydrostatic and uniaxial pressure along \textbf{c} axis, the volume
change here is significantly smaller, namely an expansion by about
0.9\%. These features are not observed in BaFe$_2$As$_2$ when
we compress along the diagonal of the $ab$-plane up to pressures
of 10 GPa.%Additionally, at a strain of 2~GPa % is 0.67~GPa meant here?
%{\Ca} undergoes a weak phase transition, also bearing traits
%of the orthorhombic to collapsed tetragonal phase transition observed
%under hydrostatic and $c$-uniaxial pressure conditions.

% (does this statement remain?).

% As strain is increased above 2GPa
%the $c$-axis contracts at an
%increased rate with $a$ and $b$ axes undergoing increased
%expansion. This is followed by a sharp decrease in iron's
%magnetic moments, from $1.76\mu_{\rm B}$ to $1.6\mu_{\rm B}$. Behavior of
%interlayer As-As distance, shown on Figure ?, sheds some light
%about this transition. We have previously found that when
%hydrostatic pressure or uniaxial pressure applied along $c$-axis
%results in formation of interlayer covalent bonds between As atoms
%once their distance is reduced bellow around $2.9\AA$. Resulting
%bond was found to be $2.79\AA$ in case of hydrostatic and
%$2.78\AA$ in case of uniaxial pressure along $c$-axis. Here, we
%observe reduction in interlayer As-As distance to around
%$2.83\AA$ indicating that weaker bonding took place, inhibited
%by iron's magnetic moments, maintaining stronger Fe-As bond
%($2.36\AA$ versus $2.30\AA$ we found in O-cT phase transition).

\subsection{Elastic constants in the orthorhombic phase}

Using data for the response to the uniaxial stress along \textbf{a},
\textbf{b} and \textbf{c}~\cite{Tomic2012} axes we can directly
evaluate the elastic constants $C_{ij}$ in {\Ba} and {\Ca}
corresponding to the orthorhombic deformations. We define elastic
constants to be such that 
\[
\sigma_{i}=\sum_{j}C_{ij}u_{j},
\]
 where $\sigma_{i}$ and $u_{j}$ are stress and strain tensor components
respectively, and indices $i$ and $j$ can be $xx,yy,zz$. Strains
are defined to be $u_{xx}=(a-a_{0})/a_{0}$, $u_{yy}=(b-b_{0})/b_{0}$
and $u_{zz}=(c-c_{0})/c_{0}$, where $a_{0}$, $b_{0}$ and $c_{0}$
are equilibrium unit cell dimensions ($P=0$~GPa). We first directly
obtain $S_{ij}=[C^{-1}]_{ij}$ by performing linear fits to $u_{i}(\sigma_{j})$
and $C$ is then obtained by inverting the resulting matrix. For {\Ba},
the elastic constant matrix is 
\begin{equation}
C=\left[\begin{matrix}95.2\pm4.3 &  & 20.4\pm3.4 &  & 40.8\pm4.5\\
27.3\pm4.8 &  & 130.8\pm6.1 &  & 64.0\pm7.0\\
43.7\pm4.5 &  & 47.7\pm4.6 &  & 81.0\pm5.6
\end{matrix}\right]\mathrm{GPa}\label{elconst_ba}
\end{equation}
 Utilizing Voigt and Reuss averages,~\cite{Clayton2010} defined
as 
\begin{align*}
B_{\mathrm{Voigt}} & =\frac{1}{9}(C_{11}+C_{22}+C_{33}+2(C_{12}+C_{13}+C_{23}))\\
B_{\mathrm{Reuss}} & =(S_{11}+S_{22}+S_{33}+2(S_{12}+S_{13}+S_{13}))^{-1}
\end{align*}
 it is possible to estimate the bulk modulus. Voigt and Reuss averages
yield $61.9\pm5.1\,$GPa and $69.3\pm7.5\,$GPa, respectively, which
is in good agreement with our previous estimate~\cite{Tomic2012}
and the experimental value of $59\pm2\,$GPa.~\cite{Jorgensen2010}
For {\Ca}, the elastic constant matrix is given by 
\begin{equation}
C=\left[\begin{matrix}148.7\pm18.5 &  & 45.6\pm12.3 &  & 55.5\pm12.7\\
63.9\pm21.4 &  & 182.4\pm18.4 &  & 81.2\pm17.5\\
61.4\pm14.7 &  & 63.1\pm11.4 &  & 68.8\pm11.3
\end{matrix}\right]\mathrm{GPa}\label{elconst_ca}
\end{equation}
 which results in bulk modulus of $84.3\pm14.8\,$GPa and $77.7\pm17.2\,$GPa
using Voigt and Reuss averages, respectively. Both values are in good
agreement with experimentally determined values of $82.9\pm1.4\,$GPa
~\cite{Mittal2011} and the estimate based on fits to Birch-Murnaghan
equation of state.~\cite{Tomic2012}

\section{Phenomenological Ginzburg-Landau model}

\begin{figure}
\includegraphics[width=0.35\textwidth]{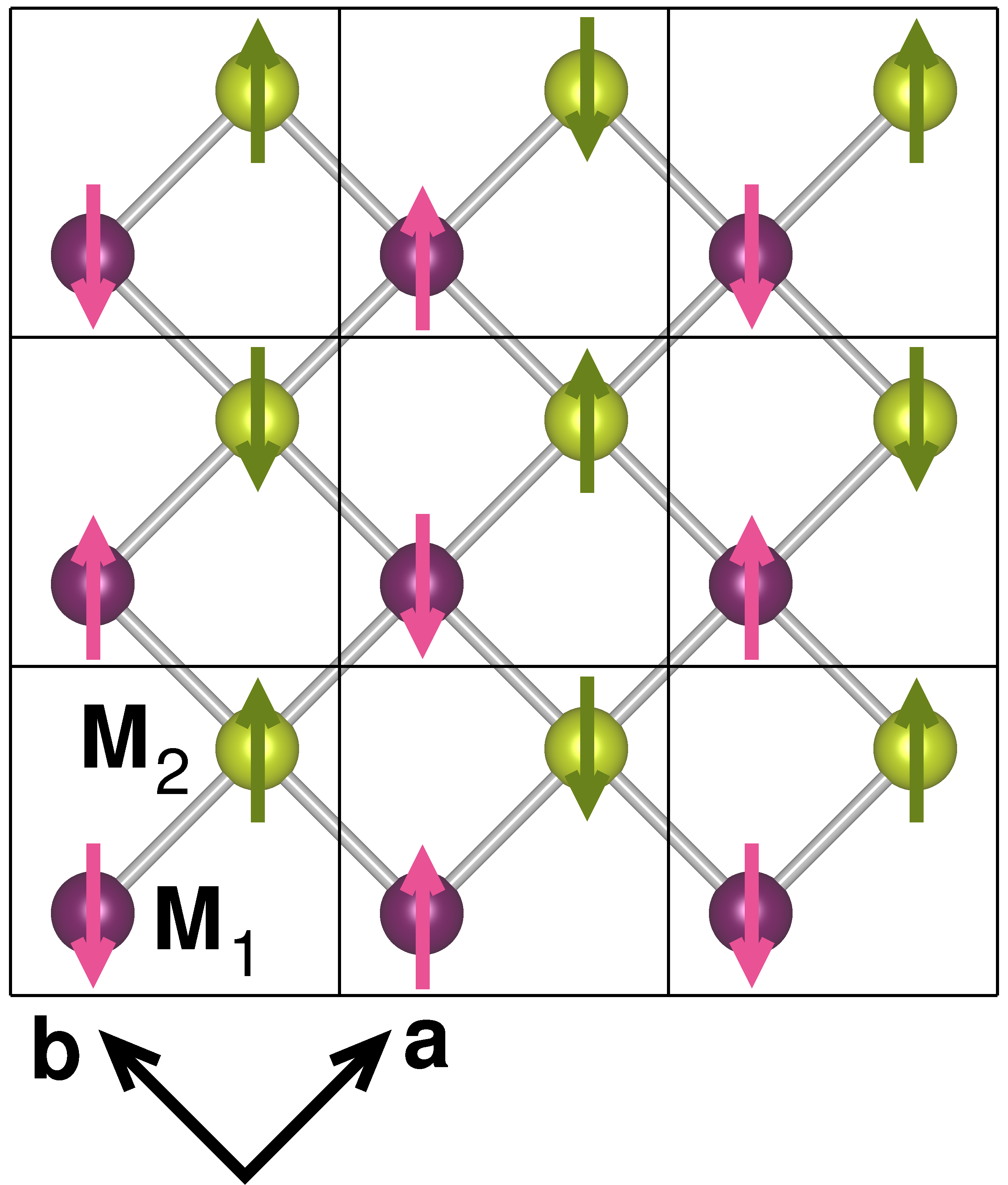} \caption{(Color online) Magnetic structure of the iron pnictides consisting
of two interpenetrating N\'eel sublattices, with magnetizations $\mathbf{M}_{1}$
and $\mathbf{M}_{2}$. \label{M1M2}}
\end{figure}

To aid the interpretation of the \emph{ab initio }results, we develop
a phenomenological magneto-elastic Ginzburg-Landau model to capture
the physics of the simultaneous sign-changing jump of the orthorhombicity
and reversal of the AFM and FM directions. As pointed out by Refs.~\onlinecite{Fang08,FernandesPRL10,Cano10},
the magnetic structure of the iron pnictides consists of two interpenetrating
N\'eel sublattices, with magnetizations $\mathbf{M}_{1}$ and $\mathbf{M}_{2}$
of equal amplitude that can point either parallel or anti-parallel
to each other (see Figure~\ref{M1M2}).

%\textbf{\emph{(should we add a figure?)}}. 
By including also the orthorhombic order parameter $\delta=\left(a-b\right)/\left(a+b\right)$,
we obtain the Ginzburg-Landau free energy: 
\begin{equation}
\begin{split}F= & \frac{a_{m}}{4}\left(M_{1}^{2}+M_{2}^{2}\right)+\frac{u_{m}}{16}\left(M_{1}^{2}+M_{2}^{2}\right)^{2}-\frac{g_{m}}{4}\left(\mathbf{M}_{1}\cdot\mathbf{M}_{2}\right)^{2}\\
 & +\frac{a_{s}}{2}\delta^{2}+\frac{u_{s}}{4}\delta^{4}+\frac{\lambda}{2}\delta\left(\mathbf{M}_{1}\cdot\mathbf{M}_{2}\right)+\sigma\delta\label{aux_{F}}
\end{split}
\end{equation}
 Here, $a_{m}\propto T-T_{N}$, $a_{s}\propto T-T_{s}$, $u_{m},u_{s}>0$,
and $g_{m}>0$. The last condition ensures that the ground state is
the striped magnetic configuration (\textit{i.e.} collinear $\mathbf{M}_{1}$
and $\mathbf{M}_{2}$). We also must have $u_{m}>g_{m}$ in order
for the magnetic free energy to be bounded. $\lambda>0$ is the magneto-elastic
coupling and $\sigma$ is the stress field conjugate to the orthorhombic
order parameter. The sign of $\lambda$ is set to describe the experimental
observation that ferromagnetic bonds are shorter than anti-ferromagnetic
bonds. Although this model does not take into account the physics
of the magnetically-driven structural transition, which comes from
fluctuations beyond the Ginzburg-Landau analysis we perform below,~\cite{FernandesPRL10}
it captures the main features of the \emph{ab initio} results.

The magnetic ground state is completely determined by the magnitude
$M=\left|\mathbf{M}_{1}\right|=\left|\mathbf{M}_{2}\right|$ and the
relative angle $\theta$ between $\mathbf{M}_{1}$ and $\mathbf{M}_{2}$.
Then, minimization of the free energy leads to three coupled equations
for $M$, $\theta$, and $\delta$: 
\begin{align}
\frac{\partial F}{\partial M} & =\left(a_{m}+\lambda\delta\cos\theta\right)M+\left(u_{m}-g_{m}\cos^{2}\theta\right)M^{3}=0\label{eq1}\\
\frac{\partial F}{\partial\delta} & =a_{s}\delta+u_{s}\delta^{3}+\frac{\lambda}{2}M^{2}\cos\theta+\sigma=0\label{eq2}\\
\frac{\partial F}{\partial\theta} & =\frac{g_{m}}{4}M^{4}\sin2\theta-\frac{\lambda}{2}M^{2}\delta\sin\theta=0\label{eq3}
\end{align}
 The last equation allows three possible solutions: $\theta=0$, $\theta=\pi$,
and $\cos\theta=\lambda\delta/\left(g_{m}M^{2}\right)$. We focus
only on the $\theta=0,\pi$ solutions, since they are the energy minimum
at zero stress. In the ordered phase, where $a_{m},a_{s}<0$, we obtain
the self-consistent equation for $\delta$: 
\begin{equation}
-\left(\left|a_{s}\right|+\frac{\lambda^{2}}{2\left(u_{m}-g_{m}\right)}\right)\delta+u_{s}\delta^{3}=-\frac{\lambda\left|a_{m}\right|\cos\theta}{2\left(u_{m}-g_{m}\right)}-\sigma\label{original_eq_delta}
\end{equation}
 For $\sigma=0$, the mean-field equations and the free energy are
invariant upon changing $\delta\rightarrow-\delta$ and $\theta\rightarrow\theta+\pi$.
Thus, we have two degenerate solutions: $\delta>0$ and anti-parallel
$\mathbf{M}_{1}$ and $\mathbf{M}_{2}$, $\theta=\pi$, (denoted hereafter
$\delta_{+}$) or $\delta<0$ and parallel $\mathbf{M}_{1}$ and $\mathbf{M}_{2}$,
$\theta=0$ (denoted hereafter $\delta_{-}$). The presence of a finite
strain $\sigma$ lifts this degeneracy. After defining: 
\begin{equation}
\begin{split}\delta_{0} & =\sqrt{\frac{\left|a_{s}\right|}{u_{s}}+\frac{\lambda^{2}}{2u_{s}\left(u_{m}-g_{m}\right)}}\\
h_{+} & =\frac{1}{u_{s}\delta_{0}^{3}}\left(\frac{\lambda\left|a_{m}\right|}{2\left(u_{m}-g_{m}\right)}-\sigma\right)\\
h_{-} & =\frac{1}{u_{s}\delta_{0}^{3}}\left(\frac{\lambda\left|a_{m}\right|}{2\left(u_{m}-g_{m}\right)}+\sigma\right)
\end{split}
\label{aux_eq_delta}
\end{equation}
 the self-consistent equations for the two solutions $\delta_{+}$
and $\delta_{-}$ become simply: 
\begin{equation}
-\left(\frac{\delta_{\pm}}{\delta_{0}}\right)+\left(\frac{\delta_{\pm}}{\delta_{0}}\right)^{3}=\pm h_{\pm}\label{eq_delta}
\end{equation}
 and we obtain analytic expressions for the two possible solutions:
\begin{equation}
\begin{split}\delta_{\pm}\left(h_{\pm}\right)= & \pm\delta_{0}\left[\left(\frac{h_{\pm}}{2}+\sqrt{\frac{h_{\pm}^{2}}{4}-\frac{1}{27}}\right)^{\frac{1}{3}}\right.\\
 & \left.+\left(\frac{h_{\pm}}{2}-\sqrt{\frac{h_{\pm}^{2}}{4}-\frac{1}{27}}\right)^{\frac{1}{3}}\right]
\end{split}
\label{sol_delta}
\end{equation}
 The interplay between the external stress field $\sigma$ and the
magneto-elastic coupling $\lambda$ becomes evident in Eqs. (\ref{aux_eq_delta})-(\ref{sol_delta}).
For $\sigma=0$, $\lambda$ acts as an external field of the same
magnitude for both the $\delta_{+}$ and $\delta_{-}$ solutions,
\textit{i.e.} it gives rise to non-zero $h_{+}=h_{-}$ in the equations
of state (\ref{eq_delta}), making these two solutions degenerate.
Now, consider that for $\sigma=0$ the system chooses the minimum
$\delta_{+}$ (\textit{i.e.} $\delta>0$ and $\theta=\pi$). By increasing
the external stress to a small value $\sigma>0$, the effective field
$h_{+}$ is suppressed, whereas the field $h_{-}$ is enhanced. Although
the solution $\delta_{-}$ (\textit{i.e.} $\delta<0$ and $\theta=0$)
has a lower energy, the solution $\delta_{+}$ is still a local minimum,
since the effective field $h_{+}$ is still finite. This situation
persists until $\sigma$ increases to the point where the field $h_{+}$
becomes negative and large enough to make the $\delta_{+}$ solution
not a local minimum.

\begin{figure}
\begin{centering}
\includegraphics[width=0.9\columnwidth]{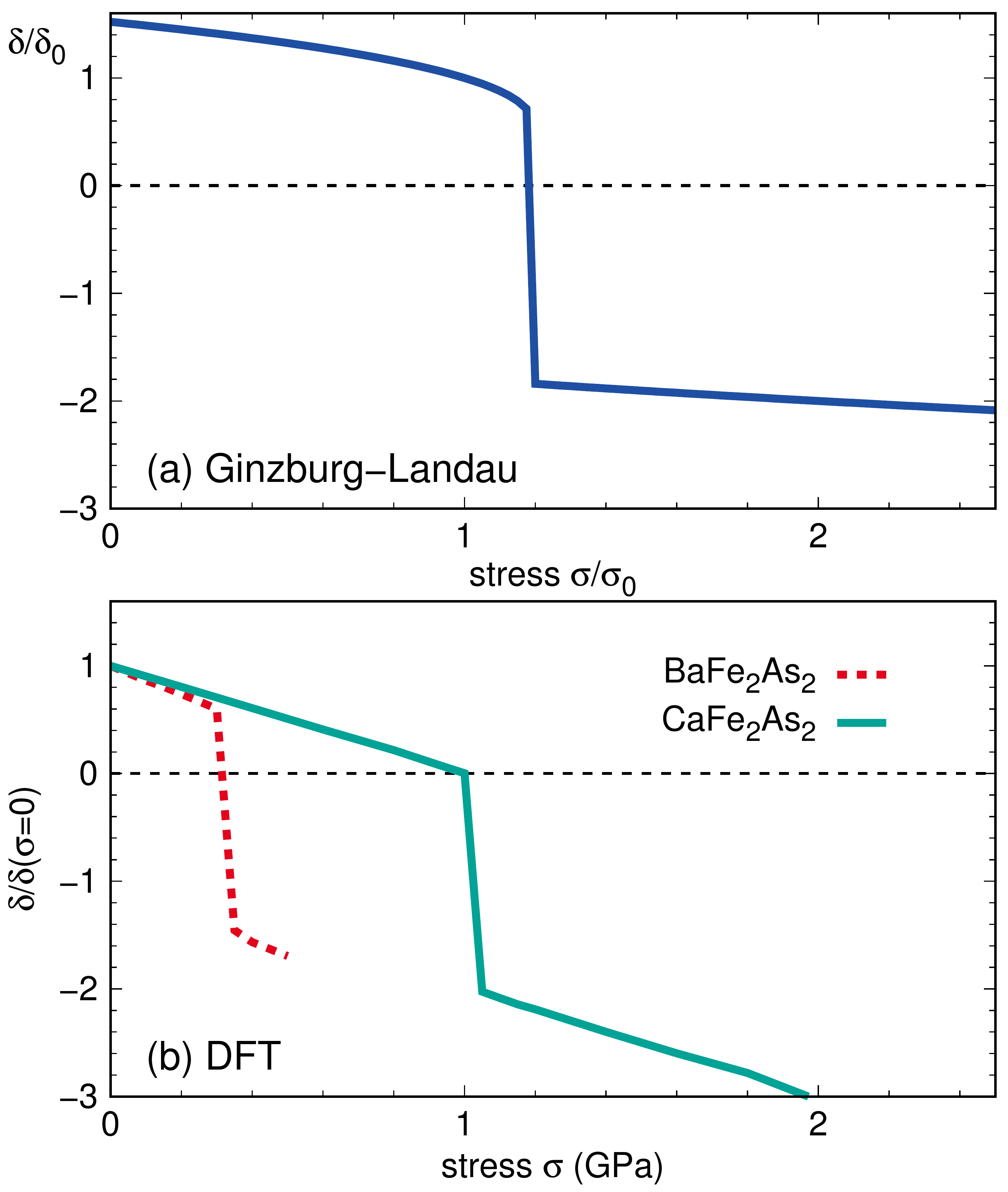} 
\par\end{centering}

\caption{(a) Orthorhombic order parameter $\delta=\frac{a-b}{a+b}$ (in
units of $\delta_{0}=\sqrt{\frac{\left|a_{s}\right|}{u_{s}}+\frac{\lambda^{2}}{2u_{s}\left(u_{m}-g_{m}\right)}}$)
as a function of the applied stress $\sigma$ (in units of $\sigma_{0}=\frac{\lambda\left|a_{m}\right|}{2\left(u_{m}-g_{m}\right)}$).
We used parameters such that $\frac{\lambda\left|a_{m}\right|}{2\left(u_{m}-g_{m}\right)u_{s}\delta_{0}^{3}}=2$.
The jump happens when the $\delta>0$ solution is no longer a local
minimum, and is accompanied by a reversal of the angle between the
two sublattice-magnetizations $\mathbf{M}_{1}$ and $\mathbf{M}_{2}$,
\textit{i.e.} a reversal of the AFM and FM directions. (b) DFT results
for the strain-dependent orthorhombic order parameter $\delta\equiv\frac{a-b}{a+b}$.
The blue curve is for {\Ba} and the red curve, for {\Ca}. \label{fig_GL}}
\end{figure}

In particular, to determine when the $\delta_{+}$ solution ceases
to be a local minimum, we analyze when one of the eigenvalues of the
Hessian matrix $\left(\partial^{2}F/\partial q_{i}\partial q_{j}\right)$
becomes negative (with generalized coordinates $q_{i}=M,\:\delta,\:\theta$).
The three eigenvalues $\mu_{i}$ are given by: 
\begin{equation}
\begin{split}\mu_{\pm} & =\frac{1}{2}\left[a_{s}+3u_{s}\delta^{2}+2M^{2}\left(u_{m}-g_{m}\right)\right]\\
 & \quad\pm\frac{1}{2}\sqrt{\left[a_{s}+3u_{s}\delta^{2}-2M^{2}\left(u_{m}-g_{m}\right)\right]^{2}+4\lambda^{2}M^{2}}\\
\mu_{0} & =\frac{M^{2}}{2}\left(g_{m}M^{2}-\lambda\delta\cos\theta\right)\label{eigenvalues}
\end{split}
\end{equation}
 For the $\delta_{+}$ ($\delta>0$, $\theta=\pi$) solution, the
only eigenvalue that can become negative with increasing $\sigma$
is $\mu_{-}$. We find that this happens when the condition $\frac{\delta_{+}}{\delta_{0}}=-\frac{3}{2}h_{+}$
is met, corresponding to an effective field $h_{+}=-\frac{2}{3\sqrt{3}}$,
\textit{i.e.} to the critical stress: 
\begin{equation}
\begin{split}\sigma_{c}= & \frac{\lambda\left|a_{m}\right|}{2\left(u_{m}-g_{m}\right)}\\
 & +\frac{2u_{s}}{3\sqrt{3}}\left(\frac{\left|a_{s}\right|}{u_{s}}+\frac{\lambda^{2}}{2u_{s}\left(u_{m}-g_{m}\right)}\right)^{3/2}
\end{split}
\label{sigma_c}
\end{equation}
 At $\sigma=\sigma_{c}$, the solution $\delta>0$, $\theta=\pi$
is not a local minimum any longer and the system jumps to the new
minimum with $\delta<0$, $\theta=0$, where not only the sign of
the orthorhombicity is reversed, but also the angle between the magnetizations
of the two sublattices (\textit{i.e.} the AFM and FM directions).
This behavior is shown in Figure~\ref{fig_GL} for a particular set
of parameters.

To compare with the DFT results, we performed a slight modification
with respect to the calculations presented in the previous section.
To ensure that the external stress couples mainly to the orthorhombic
mode $\delta$ and not to the longitudinal elastic mode $\epsilon$,
such that it does not change the volume of the system, we simultaneously
applied positive (compressive) pressure along \textbf{a }and equal-amplitude
negative (tensile) pressure along \textbf{b}. By doing this, we avoid
terms such as $M^{2}\epsilon$ in the free energy, rendering the comparison
between the \emph{ab initio }and the Ginzburg-Landau results more
meaningful.

The \emph{ab initio} obtained behavior of $\delta$ as a function of
$\sigma$, defined in the way described above, is shown also in Figure~\ref{fig_GL}.
% \textbf{\emph{(maybe you want to make the figure look
%more like the other ones?)}}.
We find a qualitative agreement with the Ginzburg-Landau results,
showing that the external stress indeed competes with the magneto-elastic
coupling, helping the system to overcome the energy barrier between
the $\delta_{+}$ ($\delta>0$, $\theta=\pi$) and $\delta_{-}$ ($\delta<0$,
$\theta=0$) solutions. A quantitative comparison becomes difficult
because the DFT calculations are performed deep in the ordered phase,
where higher order terms in the Ginzburg-Landau expansion become more
important. Furthermore, it is also possible that some of the magnetic
parameters ($a_{m}$, $u_{m}$ and $g_{m}$) have themselves some
implicit pressure dependence in this regime. Nevertheless, we can
use Eq. (\ref{sigma_c}) as a benchmark to discuss differences in
the {\Ba} and {\Ca} compounds. Clearly, Eq. (\ref{sigma_c})
shows that $\sigma_{c}$ increases with increasing magneto-elastic
coupling. Therefore, the fact that $\sigma_{c}$ is three times larger
for {\Ca} than for {\Ba} suggests that, all other parameters
being equal, the magneto-elastic coupling is larger in {\Ca} than
in {\Ba}. This may have important impact on the coupled magnetic
and structural transitions displayed by these compounds, as discussed
in Refs.~\onlinecite{Gorkov09,FernandesPRL10,Cano10}, and as such
deserves further investigation in the future.

\section{Discussion and Conclusions}

In this paper, we analyzed the effects of tensile and compressive
stress along \textbf{a}, \textbf{b} and \textbf{a}+\textbf{b} on {\Ba}
and {\Ca} by means of DFT calculations under constant stress conditions
with the help of the FIRE algorithm, combined with a phenomenological
Ginzburg-Landau model. Starting from the low-temperature magnetically
ordered orthorhombic phase, we found in the pressure range between
$-2$~GPa and $2$~GPa no real structural phase transitions in both
systems except for a pronounced orthorhombicity jump accompanied by
a 90 degree rotation of the magnetic order. FM and AFM directions
are interchanged, as are the orbital occupations $d_{xz}$ and $d_{yz}$.
This inversion of axes is a direct consequence of the interplay between
the intrinsic magneto-elastic coupling and the applied stress, as
revealed by our Ginzburg-Landau analysis. The proportionality between
the critical stress where this inversion happens and the value of
the magneto-elastic coupling suggests that in {\Ca} the magnetic
and structural degrees of freedom are more strongly coupled than in
{\Ba}, which may be related to the differences observed in their
magnetic and structural transitions.~\cite{Gorkov09} We also point
out that the estimates for the bulk moduli of {\Ba} and {\Ca}
derived from our \emph{ab initio }results are in good agreement with
the experimental measurements.

Our calculations also provide important insight on the impact of
uniaxial stress on the magnetic properties of the pnictides. Fig.
\ref{ba122moments} shows that the magnetic moment at zero temperature
always decreases (increases) with compressive (tensile) stress,
regardless of the axis that is perturbed. Unlike the jump in the
orthorhombicity and the reversal of the FM and AFM directions, this is
a consequence not of the magneto-elastic coupling, but of the changes
in the pnictogen height promoted by the uniaxial stress. This is an
important prediction of our first-principle calculations that can be
tested experimentally.  Interestingly, recent neutron diffraction
experiments~\cite{Dhital2012} on {\Ba} observed that upon application
of compressive stress along the \textbf{b} axis, the magnetic moment
is suppressed from $1.04\mu_{B}$ to $0.87\mu_{B}$. Given the small
values of applied pressure, it could be that this suppression is due
to a reduction of the volume fraction of the domains whose moments are
oriented out of the scattering plane, as pointed out by the authors of
Ref.~\onlinecite{Dhital2012}.  Nevertheless, in view of our current
results, it would be interesting to either apply higher pressures to
completely detwin the samples at low temperatures or to apply tensile
stress to make a comparison with the case of compressive stress. We
note that Ref.~\onlinecite{Dhital2012} also found an enhancement of
the magnetic transition temperature $T_{N}$ in the same detwinned
samples. Phenomenological models \cite{Cano10,Hu2012,Kuo12} attribute
this effect to changes in the magnetic fluctuation spectrum of the
paramagnetic phase promoted by the uniaxial stress. In this regard, it
would be interesting in future \emph{ab initio} studies to
systematically investigate the changes in the nesting feature of the
Fermi surface (Fig.~\ref{ba_fs}) as function of the uniaxial stress -
specifically, changes in the $\left(\pi,\pi\right)$ susceptibility
peak.

Finally, we comment on the impact of our results to the understanding
of the detwinning mechanism of iron pnictide compounds. In the tetragonal
phase, rather small uniaxial stress $P<10$ MPa is enough to completely
detwin the sample, giving rise to a single domain.~\cite{Blomberg11,Dhital2012,Fisher12}
This can be understood as fluctuations above the structural transition
temperature giving rise to long-range order in the presence of a symmetry-breaking
field.~\cite{FernandesPRL10} The situation is however very different
deep in the orthorhombic phase, where twin domains are already formed.
Experimentally, it is known that larger pressures are necessary to
completely detwin the system in this case,~\cite{Blomberg11,Fisher12}
although specific values have not been reported, to our knowledge.
One possible detwinning mechanism is the reversal of the orthorhombicity
of one domain type, while the domain walls remain pinned. This corresponds
precisely to the situation studied here, where the orthorhombicity
jumps at a certain critical uniaxial pressure. Our \emph{ab initio
}results show that such a critical pressure for {\Ba} would be
around $200$ MPa -- one order of magnitude larger than the pressure
values necessary to detwin the sample in the tetragonal phase.

Of course, other mechanisms can also give rise to detwinning in the
ordered phase, such as domain wall motion. Therefore, we propose controlled
detwinning experiments at low temperatures in {\Ba} to measure
the critical pressure necessary to form a single domain. Values comparable
to the ones discussed here would be a strong indication for reversal
of the order parameter inside fixed domains. Which mechanism is at
play in the iron pnictides may have important consequences for the
understanding of the impact of the external stress on the anisotropic
properties measured in detwinned samples -- particularly the in-plane
resistivity anisotropy,~\cite{Fisher12,Blomberg2012,FernandesPRL10}
which is likely affected by domain wall scattering.~\cite{Mazin09}

\appendix
%dummy comment inserted by tex2lyx to ensure that this paragraph is not empty

\section{Modification of the FIRE algorithm}

Within the FIRE~\cite{Bitzek2006} algorithm, the energy minimization
is achieved by moving the system's position $\bm{r}$ according to
the following equation of motion: 
\begin{equation}
\dot{\bm{v}}(t)=\frac{\bm{F}(t)}{m}-\gamma(t)|\bm{v}(t)|\left[\bm{e}_{\bm{v}(t)}-\bm{e}_{\bm{F}(t)}\right]\,,\label{fire_eq}
\end{equation}
 where $\bm{e}_{\bm{x}}$ denotes the unit vector along $\bm{x}$,
with $\bm{x}=\bm{v}(t),\bm{F}(t)$, $t$ is time and $\gamma(t)$
is a time-dependent friction parameter which assures that the system
is moving down the energy hypersurface in an optimal manner as long
as the power $P(t)=\bm{F}(t)\cdot\bm{v}(t)$ is positive. If $P(t)$
becomes negative, the procedure is stopped and relaxation is reinitialized
in the direction of the steepest descent.

In order to use FIRE for relaxation of periodic systems, we change
from the configuration space of $3N$ atomic Cartesian coordinates
$\bm{r}^{\alpha}$, $\alpha=1\dots N$, to an extended system of $3N+9$
coordinates $\tilde{\bm{r}}^{\alpha}=(\bm{s}^{\alpha},\mat{h})$,
consisting of lattice vectors which are contained in columns of the
$3\times3$ matrix $\mat{h}=\left(h_{ij}\right)$, and fractional
atomic positions $\bm{s}^{\alpha}$ within the unit cell, where Cartesian
and fractional positions are related by $\bm{r}^{\alpha}=\mat{h}\bm{s}^{\alpha}$.

When \eqref{fire_eq} is rewritten in terms of coordinates $\tilde{\bm{r}}^{\alpha}$,
one just needs to find the appropriate expression for forces $\tilde{\bm{F}}^{\alpha}$,
that is, derivatives of energy with respect to the coordinates $\tilde{\bm{r}}^{\alpha}$.
Since stress and strain tensors, $\hat{\sigma}$ and $\mat{u}$, can
be defined through 
\[
\mat{\sigma}=-\frac{1}{V}\frac{\partial E}{\partial\mat{u}},\quad\mat{H}=\Big(\mat{I}+\mat{u}\Big)\mat{h},
\]
 where $\mat{I}$ is the identity matrix and $\mat{H}$ is the lattice
matrix after an infinitesimal deformation, it is easy to show that
\begin{equation}
\begin{split}\frac{\partial E}{\partial\bm{s}^{\alpha}} & =\sum_{\beta}\frac{\partial E}{\partial\bm{r}^{\beta}}\frac{\partial\bm{r}^{\beta}}{\partial\bm{s^{\alpha}}}=\frac{\partial E}{\partial\bm{r}^{\alpha}}\mat{h}^{T}=\bm{F}^{\alpha}\mat{h}^{T},\\
\frac{\partial E}{\partial\mat{h}} & =\frac{\partial E}{\partial\mat{u}}\frac{\partial\mat{u}}{\partial\mat{h}}=\frac{\partial E}{\partial\mat{u}}\Big(\mat{h}^{T}\Big)^{-1}=-V\mat{\sigma}\Big(\mat{h}^{T}\Big)^{-1},
\end{split}
\end{equation}
 so that forces in the extended coordinates are given by 
\begin{equation}
\tilde{\bm{F}}^{\alpha}=\bigg[\bm{F}^{\alpha}\mat{h}^{T},-V\mat{\sigma}\Big(\mat{h}^{T}\Big)^{-1}\bigg].\label{forces}
\end{equation}
 Forces $\bm{F}^{\alpha}$ and stresses $\mat{\sigma}$ are obtained
from the electronic structure code, and are inserted directly into
Eq.~\eqref{forces}, taking into account that $\mat{\sigma}=\mat{\sigma}^{ext}-\mat{\sigma}^{int}$,
that is, total stress is the sum of internal and external stresses
applied to the surface of the unit cell.

%\begin{center}
%\includegraphics[width=0.43\textwidth]{ba122_freeenergy}
%\caption{(Color online) Behaviour of contributions to the free energy
%  near the stresses (marked by dotted and dashed lines) where
%  ferromagnetic and antiferromagnetic directions are exchanged. Lines
%  joining the calculated points are a guide for the eye.}
%\label{ba122freeenergy}
%\end{center}
%\end{figure}

%Putting our estimates for the known parameters into \eqref{fren_cont} we
%determine for {\Ba} $g=2.2\,{\rm GPa}/\mu_{\rm B}^2$
%and $B=4.9\cdot 10^5$GPa, and for {\Ca} $g=2.7\,{\rm GPa}/\mu_{\rm B}^2$
%and $B=1.6\cdot 10^5$GPa.

% For the evaluation of the bulk modulus we have used generalized
% Birch-Murnaghan equation of state (EOS), as derived in \cite{Sata2002}.
% This form of EOS uses state at an arbitrary pressure as a reference
% state, in constrast to a regular EOS, which uses zero-pressure state
% as a reference. If a system, at a pressure $p_r$, has a volume $V_r$,
% bulk modulus $B_r$, and a bulk modulus derivative $B'_r$, then pressure
% at a volume $V$ is given by
% \\
% \begin{align}\label{bmeos}
% \begin{split}
% p &= \left[p_r-\frac{1}{2}(3B_r-5p_r)f\right. \\
% &{\qquad} +\left.\frac{9}{8}B_r\left(B'_r-4+
% \frac{35p_r}{9B_r}\right)f^2\right](1-f)^{5/2},
% \end{split}
% \end{align}
% \\
% where
% \\
% \begin{equation*}
% f = \left[1-\left(\frac{V_r}{V}\right)^{2/3}\right].
% \end{equation*}
% \\
% Fitting this equation to every $(p, V)$ point of computed data
% as a reference state, one can obtain $B_r$ and $B'_r$.

\textbf{Acknowledgments.-} We would like to thank Igor I. Mazin for
useful discussions and the Deutsche Forschungsgemeinschaft for financial
support through grant SPP 1458, the Helmholtz Association for support
through HA216/EMMI and the centre for scientific computing (CSC, LOEWE-CSC)
in Frankfurt for computing facilities. This research was supported
in part (RV) by the National Science Foundation under Grant No. NSF
PHY11-25915.

%reference priority classified as top, med, low

%low
%1


\begin{thebibliography}{10}
\bibitem{Kamihara2008} Y. Kamihara, T, Watanabe, M, Hirano, H. Hosono,
J. Am. Chem. Soc. \textbf{130}, 3296 (2008).


%2


\bibitem{Chu2010} J. H. Chu, J. G. Analytis, K De Greeve, P. L. McMahon,
Z. Islam, Y. Yamamoto, I. R. Fisher, Science \textbf{329}, 824 (2010).


%3


\bibitem{Tanatar2010} M. A. Tanatar, E. C. Bloomberg, A. Kreyssig,
M. G. Kim, N. Ni, A. Thaler, S. L. Bud'ko, P. C. Canfield, A. I. Goldman,
I. I. Mazin, R. Prozorov, Phys. Rev. B \textbf{81}, 184508 (2010).


%4


\bibitem{Kuo2011} H.-H. Kuo, J. H. Chu, S. C. Riggs, L. Yu, P. L.
MacMahon, K. De Greeve, Y. Yamamoto, J. G. Analytis, I. R. Fisher,
Phys. Rev. B \textbf{84}, 054540 (2011).


%5


\bibitem{Liang2011} T. Liang, M. Nakajima, K. Kihou, Y. Tomioka,
T. Ito, C. H. Lee, H. Kito, A. Iyo, H. Eisaki, T. Kakeshita, S. Uchida,
J. Phys. Chem. Solids \textbf{72}, 418 (2011).


%6


\bibitem{Dhital2012} C. Dhital, Z. Yamani, W. Tian, J. Zeretsky,
A. S. Sefat, Z. Wang, R. J. Birgeneau, S. D. Wilson, Phys. Rev. Lett.
\textbf{108}, 087001 (2012).


%7


\bibitem{Blomberg2012} E. C. Blomberg, A. Kreyssig, M. A. Tanatar,
R. M. Fernandes, M. G. Kim, A. Thaler, J. Schmalian, S. L. Bud'ko,
P. C. Canfield, A. I. Goldman, R. Prozorov, Phys. Rev. B \textbf{85},
144509 (2012).


%8


\bibitem{Fisher12} J.-H. Chu, H.-H. Kuo, J. G. Analytis, and I. R.
Fisher, Science \textbf{337}, 710 (2012).

\bibitem{Fang08} C. Fang, H. Yao, W.-F. Tsai, J. Hu, and S. A. Kivelson,
Phys. Rev. B \textbf{77,} 224509 (2008).

\bibitem{Xu08} C. Xu, M. M\"uller, and S. Sachdev, Phys. Rev. B \textbf{78},
020501(R) (2008).

\bibitem{FernandesPRL10} R. M. Fernandes, L. H. VanBebber, S. Bhattacharya,
P. Chandra, V. Keppens, D. Mandrus, M. A. McGuire, B. C. Sales, A.
S. Sefat, and J. Schmalian, Phys. Rev. Lett. \textbf{105}, 157003
(2010).

\bibitem{Fernandes11} R. M. Fernandes, E. Abrahams, and J. Schmalian,
Phys. Rev. Lett. \textbf{107}, 217002 (2011).

\bibitem{Blomberg11} E. C. Blomberg, M. A. Tanatar, A. Kreyssig,
N. Ni, A. Thaler, R. Hu, S. L. Bud'ko, P. C. Canfield, A. I. Goldman,
and R. Prozorov, Phys. Rev. B 83, 134505 (2011).

\bibitem{Kuo12} H-.H. Kuo, J. G. Analytis, J.-H. Chu, R. M. Fernandes,
J. Schmalian, and I. R. Fisher, Phys. Rev. B \textbf{86}, 134507 (2012).

\bibitem{Wang2012} Q. Y. Wang, Z. Li, W. H. Zhang, Z. C. Zhang, J.
S. Zhang, W. Li, H. Ding, Y. B. Ou, P. Deng, K. Chang, J. Wen, C.
L. Song, K. He, J. F. Jia, S. H. Ji, Y. Wang, L. Wang, X. Chen, X.
Ma, Q.K. Xue, Chin. Phys. Lett. \textbf{29}, 037402 (2012).


%10


\bibitem{Cano2012} A. Cano and I. Paul, Phys. Rev. B \textbf{85},
155133 (2012).


%11


\bibitem{Kresse1993} G. Kresse and J. Hafner, Phys. Rev. B \textbf{47},
558 (1993); G. Kresse and J. Furthm\"uller, ibid. \textbf{54}, 11169
(1996); Comput. Mater. Sci. \textbf{6}, 15 (1996). %12


\bibitem{Bloechl1994} P. E. Bl\"ochl, Phys. Rev. B \textbf{50}, 17953
(1994); G. Kresse and D. Joubert, ibid. \textbf{59}, 1758 (1999).


%16


\bibitem{Bitzek2006} E. Bitzek , P. Koskinen, F. G{\"a}hler, M. Moseler,
P. Gumbsch, Phys. Rev. Lett. \textbf{97}, 170201 (2006).


%14


\bibitem{Tomic2012} M. Tomi\'{c}, R. Valent{\'\i}, H. O. Jeschke,
Phys. Rev. B \textbf{85}, 094105 (2012).


%9


\bibitem{Zhang2009} Y.-Z. Zhang, H. C. Kandpal, I. Opahle, H. O.
Jeschke, and R. Valent{\'\i}, Phys. Rev. B \textbf{80}, 094530 (2009).

\bibitem{Colonna2011a} N. Colonna, G. Profeta, A. Continenza, and
S. Massidda, Phys. Rev. B \textbf{83}, 094529 (2011).

\bibitem{Colonna2011b} N. Colonna, G. Profeta, and A. Continenza,
Phys. Rev. B \textbf{83}, 224526 (2011).

\bibitem{Aichhorn2009} M. Aichhorn, L. Pourovskii, V. Vildosola,
M. Ferrero, O. Parcollet, T. Miyake, A. Georges, and S. Biermann Phys.
Rev. B \textbf{80}, 085101 (2009).

\bibitem{Yin2011} Z. P. Yin, K. Haule, and G. Kotliar, Nature Mater.
\textbf{10}, 932 (2011).

\bibitem{Aichhorn2011} M. Aichhorn, L. Pourovskii, and A. Georges,
Phys. Rev. B \textbf{84}, 054529 (2011).

\bibitem{Ferber2012a} J. Ferber, K. Foyevtsova, R. Valent{\'\i},
and H. O. Jeschke, Phys. Rev. B \textbf{85}, 094505 (2012).

\bibitem{Ferber2012b} J. Ferber, H.O. Jeschke, and R. Valent{\'\i},
Phys. Rev. Lett. \textbf{109}, 236403 (2012). %17

\bibitem{Werner2012} 
    P. Werner, M. Casula, T. Miyake, F. Aryasetiawan, A.J. Millis, 
and S. Biermann,
Nature Physics, {\bf 8}, 331 (2012).
\bibitem{Hu2012} J. Hu, C. Setty, and S. Kivelson, Phys. Rev. B \textbf{85},
100507(R) (2012).

\bibitem{Kreyssig2008} A. Kreyssig, M. A. Green, Y. Lee, G D. Samolyuk,
P. Zajdel, J. W. Lynn, S. L. Bud{'}ko, M. S. Torikachvili, N. Ni,
S. Nandi, J. B. Le{\~a}o, S. J. Poulton, D. N. Argyriou, B. N. Harmon,
R. J. McQueeney, Phys. Rev. B \textbf{78}, 184517 (2008).

\bibitem{Clayton2010} J. D. Clayton, \textit{Nonlinear Mechanics
of Crystals}, Springer (2010).


%15


\bibitem{Jorgensen2010} J. E. J{\"o}rgensen, T. C. Hansen, Eur. Phys.
J. B \textbf{78}, 411 (2010).

\bibitem{Mittal2011} R. Mittal, S. K. Mishra, S. L. Chaplot, S. V.
Ovsyannikov, E. Greenberg, D. M. Trots, L. Dubrovinsky, Y. Su, Th.
Brueckel, S. Matsuishi, H. Hosono, G. Garbarino, Phys. Rev. B \textbf{83},
054503 (2011).

\bibitem{Boeri} O. K. Andersen, L. Boeri, Ann. Phys. \textbf{523},
8 (2011).

\bibitem{Cano10} A. Cano, M. Civelli, I. Eremin and I. Paul, Phys.
Rev. B \textbf{82}, 020408(R) (2010).

\bibitem{Gorkov09} V. Barzykin and L. P. Gor'kov, Phys. Rev. B \textbf{79},
134510 (2009).

\bibitem{Mazin09} I. I. Mazin and M. D. Johannes, Nature Phys. \textbf{5},
141 (2009). \end{thebibliography}
\end{document}